    \documentclass[aps,prl,twocolumn,groupedaddress,showpacs]{revtex4}
    \usepackage{graphicx}

\begin{document}


    \title{Electron correlation effects on the dielectric function of liquid 
metals }

    \author{P. Giura$^{1}$, R. Angelini$^{1}$, C. A. Burns$^{2}$, 
G. Monaco$^{1}$ and F. Sette$^{1}$.}

    \affiliation{ $^{1}$ European Synchrotron Radiation Facility. 
             B.P. 220 F-38043 Grenoble, Cedex France.\\ 
             $^2$ Department of Physics, Western Michigan University, 
    Kalamazoo, Michigan 49008}

    \date{\today}

    \begin{abstract}
    The acoustic excitations of the expanded metal solutions Li-NH$_3$ 
    have been measured by inelastic X-ray scattering as a 
    function of the electron density by changing the Li concentration. 
    The dielectric functions of these model metals with very low electron density
    have been derived from the high frequency sound velocity using the 
    one component plasma approach corrected for screening. 
    Their values, when combined with those from other metals, suggest that the
    electron correlation induced departure of the dielectric function from the
    random phase approximation follows a universal behavior whose main parameter
    is the electron density.

    \end{abstract}


    \maketitle

    The study of the dynamical properties of interacting electrons 
    is an important topic in solid state physics as shown by the great 
interest devoted to the understanding of phenomena as superconductivity
    and giant magneto-resistance.
    One way to obtain information on the dielectric response of a system of 
interacting electrons is to study the acoustic excitations of liquid metals, 
    since the speed of sound in these systems is related to the screening action 
of the electrons 
    on the ionic motion ~\cite{AshcroftMermin}. In fact, in a simple Jellium model with
the electrons screening the Coulomb forces between the ions, the speed 
    of sound, $c$, can be expressed as:
    \begin{equation}
    c = {{\Omega_{pi}} \over { Q \sqrt{\epsilon(Q)} }}.
    \label{cs}
    \end{equation}
    Here Q is the wave number, $\epsilon(Q)$ is the total dielectric function
    and $\Omega_{pi}$ is the ionic plasma frequency given by: 
    $\Omega^2_{pi}=(4 \pi n_i(Ze)^2)/M$, where $M$, $Ze$ and $n_i$ are the mass, 
    the charge and the number density of the ions, 
    respectively. In this relation all the details of the electron dynamics are 
    hidden in the dielectric function, $\epsilon(Q)$, that is purely electronic only if
    the ions contribution can be neglected - this is the case at high frequencies and 
    $Q$-transfers, where the atomic response of the liquid is totally elastic 
and one measures the 
    {\it infinite}-frequency value of the sound velocity, $c$. 
    The strength of Eq.~\ref{cs} is to provide a direct link between an electronic property as
  $\epsilon(Q)$ and a ionic one as $c$.
     Therefore, the measurements of $c$ offers the opportunity to check experimentally 
the validity 
    of theoretical approaches for $\epsilon(Q)$ as, for example, the
    Random Phase Approximation (RPA), where~\cite{AshcroftMermin}:
    \begin{equation}
    \epsilon_{RPA}(Q) = 1 + {{4 m^2 e^2 v_F} \over {\pi \hbar^3 Q^2}}=1+{{Q_{RPA}^2} \over {Q^2}}.
    \label{epsilon}
    \end{equation}
    Here, $m$ is the electron mass, $v_F$ the Fermi velocity and $Q_{RPA}$ is a 
    measure of the inverse distance over which the self-consistent electric 
    potential associated with the ions penetrates into the electron gas.
    This approximation is exact in the limit of high electron density, 
    or $r_s$$\to$0. The parameter $r_s=[3/(4\pi n_e)]^{1/3}a_0^{-1}$, with $a_0$ the Bohr
    radius and $n_e$ the electron number density, is the ratio between the Coulomb 
    and the kinetic energies of the electrons. The  $r_s$$\to$0 limit corresponds, therefore, to 
    the weak coupling limit. As a matter of fact, the RPA works well in metals with a low 
value of $r_s$ and with no unfilled {\it d} or {\it f} shells, as shown in the case 
    of Hg and Pb where $r_s$ is $\approx$ 3 ~\cite{BovePRL01}. 
    In metals with higher values of $r_s$ the RPA clearly fails as recently 
    reported for the case of the Li-NH$_3$ solution at the limit of solubility for which $r_s$ 
    is $\approx$ 7 ~\cite{BurnsPRL99,BurnsPRL02,SacchettiPRB03} and for heavy 
alkali metals ~\cite{BovePRB03}.
    It is believed that the failure of the RPA at large $r_s$ is mainly due to 
    electronic correlation effects. In this context the Li-NH$_3$ solutions provide an 
interesting model system to investigate the departure of the $\epsilon(Q)$ from the 
RPA predictions as a function of the electron density  
- here the $r_s$ value can be widely varied by changing the
    Li concentration from the saturated solution value $r_s$=7.4 to the metal-insulator 
transition value $\approx$ 12. 

In this Letter we report the experimental determination of the high frequency velocity 
of sound in Li-NH$_3$ solutions and the derivation of the corresponding dielectric function 
using Eq~\ref{cs}.
This provides a context to 
    investigate the departure of the $\epsilon(Q)$ from the RPA predictions as a function 
of the electron density - here the $r_s$ value has been changed between 7.4 and 10.4.
    These results, when complemented by literature data on liquid metals with lower 
    values of $r_s$ ~\cite{BovePRL01,CopleyPRA74,CopleyPRL74,BovePRL00,
    ScopignoPRE02,ScopignoEL00,SoderstromJPF80,ScopignoPRE00,CabrilloPRL02,
    DornerPRA92}, provide i) the $r_{sth}$=4.6 value as the threshold value below which 
    the RPA properly works and ii) the characteristic dependence of the electronic 
    dielectric function on the electron density at $r_s$ larger than $r_{sth}$. 
    These findings suggest the remarkable result that the electron density is the relevant 
    parameter in the description of the dielectric function even when the electron 
    response starts to be affected by complex corrections to the RPA.

    The experiment was performed at the high resolution Inelastic X-rays 
    Scattering (IXS) beamline ID16 at the European Synchrotron Radiation Facility 
    in Grenoble (France). The measurements were carried out with a total energy resolution 
of  $\sim$ 
    1.5 meV and an exchanged momentum, $Q$, resolution of 0.4 nm$^{-1}$. 
    Further details of the beamline are reported elsewhere~\cite{MasciovecchioNIMPR96}.
    Every scan took about 210 min and the final spectra were obtained by averaging
    3 to 4 scans. 
    The Q values were selected between 1 and 15 nm$^{-1}$.
    The samples were prepared using high purity Li (nominally 99.9 \%) and high 
    purity anhydrous ammonia (nominally 99.995\%). A stainless steel container, 
    with two sapphire windows (total thickness 500 $\mu$) and with 5 cm$^3$ internal 
    volume, was kept in a cryostat at T=240.0$\pm$0.2 K. 
    The initial solution was prepared at the limit of solubility of lithium 
    in ammonia (20$\pm$0.5 mpm). Successive dilutions were carried out
    \textit{in situ} in order to measure three other metallic concentrations: 
17.5$\pm$0.5 mpm, 9.3$\pm$0.2 mpm and 6.5$\pm$0.2 mpm.

\begin{figure}
    \includegraphics[width=7.5cm,height=9.5cm]{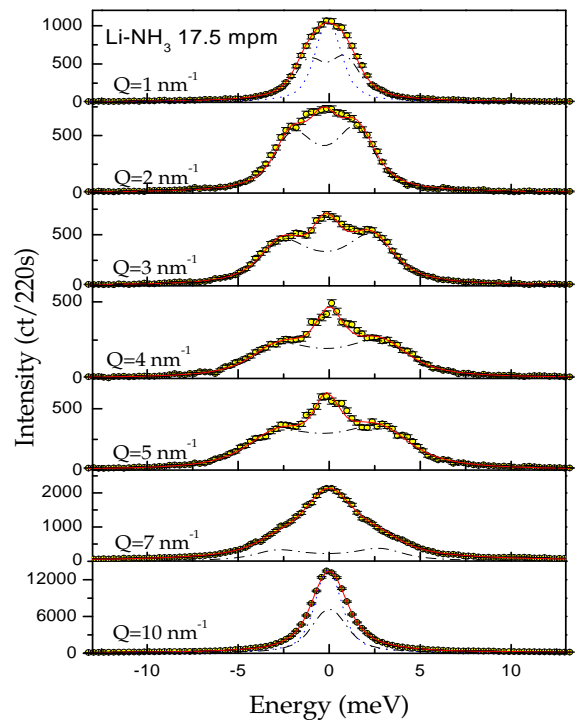}
    \caption{Selected IXS spectra for the 17.5 mpm LiNH$_3$ solution (o)
    at the indicated values of the exchanged wave vector, plotted together with 
    the total fitting function (-) corresponding to Eq.~\ref{eq5} and the 
    resolution convoluted inelastic features (dot-dashed line). At Q=1 nm $^{-1}$ 
    and Q=10 nm $^{-1}$ the experimental resolution function (dotted line), 
scaled to the maximum of the spectrum, is also reported.}
    \label{fig1} 
    \end{figure}

\begin{figure} 
    \includegraphics[width=7cm,height=8.5cm]{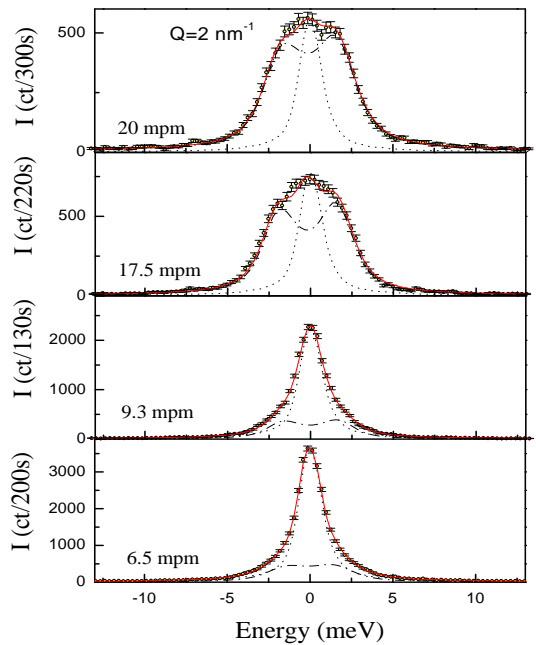}
    \caption{Selected IXS spectra at fixed wave vector transfer Q=2 nm$^{-1}$ for
    the four studied solutions. The experimental data (o) are 
    plotted together with the experimental resolution (dotted line), the total
    fitting function (-) and the inelastic resolution-convoluted contributions 
    (dot-dashed line).}
    \label{fig2} 
    \end{figure}

\begin{figure}
    \includegraphics[width=7cm,height=8.5cm]{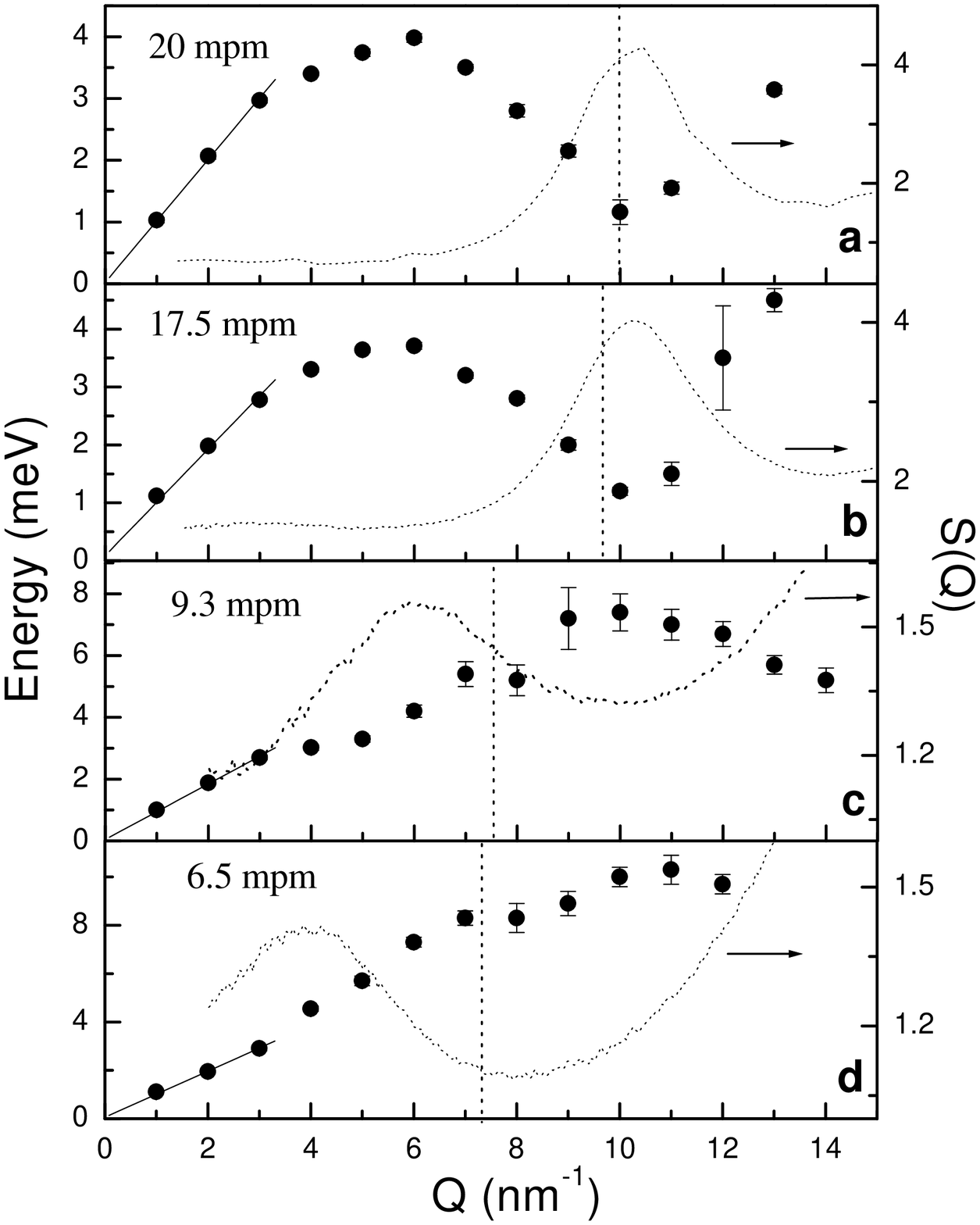}
    \caption{Dispersion curves for the acoustic excitations obtained from the fit 
    procedure discussed in the text for the Li-NH$_3$ solutions at the 
    indicated concentrations. The X-ray S(Q) spectra are also 
    reported (right axes)~\cite{S(Q)nostre}. The vertical dotted lines indicate 
    the wave vector position corresponding to 2 $K_F$. The solid lines 
    are the low Q linear fits to the dispersion curves.} 
    \label{fig3} 
    \end{figure}

    IXS spectra are reported in Fig.~\ref{fig1} for the solution at 17.5 
    mpm at a representative set of Q values. They consist of a central 
    quasielastic line and two clear inelastic components which, as shown 
    in Fig.~\ref{fig1}, disperse up to Q=7 nm$^{-1}$. 
    The dependence of the measured spectra on the Li concentration is shown 
    in Fig.~\ref{fig2} at Q=2 nm$^{-1}$ for the four studied solutions.
    The spectral line shape changes considerably on changing the Li content:
    the quasi-elastic line becomes more and more intense with respect to the 
    inelastic features as the metal to insulator transition is approached 
    (top to  bottom spectra in Fig.~\ref{fig2}). At larger $Q$ values, particularly at
    the first sharp diffraction peak ($Q_M$ $\sim$ 10 nm$^{-1}$ for the solution at 17.5 mpm,
    see Fig.~\ref{fig3}b), the spectra become narrow with a width 
    comparable to the resolution function. 

    The measured IXS spectra have been fitted to the convolution of the 
    resolution function with a model function composed of a Lorentzian curve 
    for the central line and a Damped Harmonic Oscillator (DHO) function for 
    the inelastic signal~\cite{Balucani94}:
    \begin{eqnarray}
    \nonumber
    S(Q,E)={{E/K_BT}\over{1-e^{E/K_BT}}}\Big({{I_c(Q)\Gamma_c(Q)}
    \over{\Gamma^2_c(Q)+E^2}}+ \\
    {{I(Q)\Gamma(Q)\Omega^2(Q)}\over{(\Omega^2(Q)-E^2)^2+4\Gamma^2(Q)E^2}}\Big).
    \label{eq5}
    \end{eqnarray}
    The quasielastic and inelastic spectral features are defined in terms of a 
    set of two and three parameters, respectively. Those concerning the inelastic 
    components are related to the energy, $\hbar\Omega$(Q), lifetime, 
    $\Gamma$(Q), and strength, I(Q), of the acoustic-like excitations.
    Both Fig.~\ref{fig1} and Fig.~\ref{fig2} show that the fitting function
    of Eq.~\ref{eq5} describes well the experimental spectra.
    The $Q$-dispersions of the fitting parameter $\Omega$(Q) are shown in 
    Fig.~\ref{fig3} (left axes) together with X-ray measurements of the 
    static structure factor, S(Q)~\cite{S(Q)nostre} (right axes), and an indication 
    of the location of twice the Fermi wavenumber, 2$K_F$ (dotted vertical line). 
    At low Q the acoustic excitations of the solutions show a linear 
    dispersion (solid lines in Fig.~\ref{fig3}) as it is expected for a 
    sound-like mode. 
    In the panels (a) and (b) of Fig.~\ref{fig3} the results for the two more 
    metallic solutions are reported. Here, $\Omega$(Q) shows a maximum between
    5 and 6 nm$^{-1}$ and then decreases reaching a minimum at a Q value 
    close to the first maximum, $Q_M$, in the corresponding S(Q). This is a common 
    finding in liquids which is usually referred to as the De Gennes 
    narrowing~\cite{Balucani94}. A different interpretation of this minimum 
    formulated in terms of the Q-dependence of the electronic dielectric function 
    has instead been proposed for the 20 mpm solution 
    ~\cite{BurnsPRL01,SacchettiPRB03}. 
    In fact, for Q-values of the order of $K_F$, Eq.~\ref{epsilon} must be 
    replaced by the more accurate Lindhard result~\cite{AshcroftMermin} 
    which has a singular point at Q = 2 $K_F$. This singularity would in turn 
    induce a kink 
    in the dispersion curve via Eq.~\ref{cs}. However, given the proximity of 
    2$K_F$ to $Q_M$ for both the 20 mpm and the 17.5 mpm solutions, it is hard 
    to disentangle structural and electronic effects.  
    The dispersion curves of the solutions at 9.3 mpm and 6.5 mpm reported in 
    the panels (c) and (d) of Fig.~\ref{fig3} are, from this point of view, 
    much more interesting since 2$K_F$ and $Q_M$ are here considerably different.
    As a matter of fact, these two dispersion curves both show a slight 
    bending down in the proximity of the first diffraction peak (less evident for 
    the solution at 6.5 mpm) and a small kink at Q$\sim$8 nm$^{-1}$, very close to 
    2 $K_F$. This kink suggests indeed the presence of a Kohn-like 
    instability of the system at Q $\sim$ 2 $K_F$. This result - if confirmed 
    with a better statistics and higher Q-resolution - 
    would then delineate the effects of the electron-phonon coupling in these 
    expanded liquid metals, and support the interpretation 
    proposed earlier ~\cite{BurnsPRL01,SacchettiPRB03}. 
    However, these data indicate that in any case the Kohn anomaly has only 
    a small effect on the dispersion curve, and the minimum at $Q_M$ has mainly a
    structural origin.

    Linear fits to the dispersion curves at low Q, reported in 
    Fig.~\ref{fig3}, are used to determine the high frequency sound velocity of the 
measured solutions:
     $c_{Exp}$(20 mpm)=1580$\pm$20 m/s, $c_{Exp}$(17.5 mpm)=1519$\pm$60 m/s, 
$c_{Exp}$(9.3 mpm)=1430$\pm$60 m/s and $c_{Exp}$(6.5 mpm)=1450$\pm$60 m/s ~\cite{nota1}. 
These values are compared with the RPA expectations obtained combining Eqs.
    \ref{cs} and \ref{epsilon}, {\it i.e.} $c_{RPA}=v_F\sqrt{(Zm)/(3M)}$ 
    ~\cite{AshcroftMermin}. In this expression, knowing that the Li ions form
    tetrahedral complexes with the ion surrounded by four ammonia molecules 
    ~\cite{WasseJCP99,WassePRB00}, $M$ is identified as the mass of the Li(NH$_3$)$_4$ 
complex, as already proposed in refs.~\cite{BurnsPRL01,SacchettiPRB03}. 
    In this picture, the solution at the limit of solubility (20 mpm)
    is completely made up of these complexes, while at lower Li content,
    the solutions can be viewed as a two component system where the Li(NH$_3$)$_4$
    complexes are still stable entities embedded in the uniform background of the ammonia 
molecules ~\cite{KnappJAC78}. 
The comparison between the $c_{Exp}$ and $c_{RPA}$ is presented
    in Fig. 4 as the ratio $c^2_{Exp}$ / $c^2_{RPA}$ as a function of $r_s$.
    The corresponding $\epsilon_{RPA}/\epsilon_{Exp}$ ratio is on the left axis.
    In Fig.4 we also report the results obtained
    for other liquid metals, as taken in the literature from ref.~\cite{ScopignoPRE00} for Al, 
    ref.~\cite{BovePRL01} for Hg, ref.~\cite{SoderstromJPF80} for Pb, 
    ref.~\cite{ScopignoEL00} for Li, ref.~\cite{ScopignoPRE02} for Na,
    ref.~\cite{CabrilloPRL02} for K, ref.~\cite{DornerPRA92} for Cs and
    ref.~\cite{CopleyPRL74} for Rb. 
    At low $r_s$, as expected, a very good agreement
    exists between the experimental results and the RPA predictions, confirming nicely
    that Eq.~\ref{cs} relates the infinite frequency sound velocity and the dielectric 
function.
    A clear departure from the RPA is observed in Fig. 4 at the threshold value
    $r_{sth}$=4.6, above which $c_{Exp}$ becomes sensibly larger than $c_{RPA}$.
    
    The very intriguing result of Fig. 4 is that all the existing data follow a continuous
    trend despite of the very large variety of considered systems: steric
    effects, chemical specificities, and spatial extent of the different atoms and complexes
    are somehow irrelevant for $c^2_{Exp}$ / $c^2_{RPA}$ and
    $\epsilon_{RPA}/\epsilon_{Exp}$ - the only important parameter is $r_s$. In other words, 
regardless of a number of parameters not correctly treated by the RPA, as electronic correlations and finite ionic size effects, the electron dynamics has a general dependence on the electron density.
    \begin{figure} 
    \includegraphics[width=7.5cm,height=6cm]{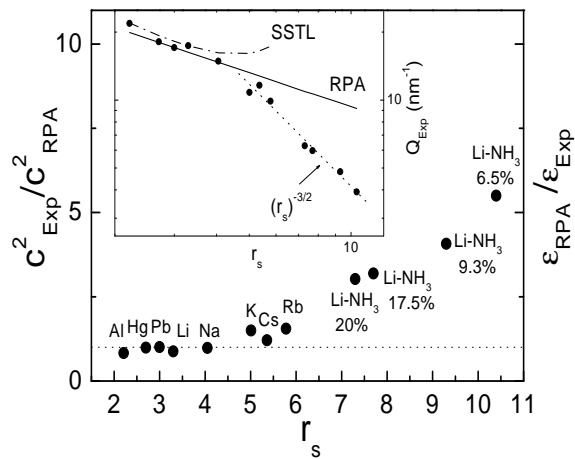}
    \caption{Ratio between the experimental and the RPA squared sound 
    velocities (left axis) as a function of $r_s$ for 
    different liquid metals~\cite{ScopignoPRE00,BovePRL01,
    SoderstromJPF80,ScopignoEL00,ScopignoPRE02,CabrilloPRL02,DornerPRA92,
    CopleyPRL74}. These data correspond via Eq.1 to the ratio between the RPA and 
    the experimental dielectric function (right axis).
    The dotted line corresponds to the unity value.
    In the insert the corresponding experimental screening wave vectors 
    (closed circles) are compared to the RPA (solid line) and to the SSLT 
    theory (dot dashed line). The dotted line emphasizes the phenomenological 
    $r_s^{-3/2}$ law obeyed by metals for $r_s\ge$4.6.}
    \label{fig4} 
    \end{figure}

    The results reported in Fig. 4 are further emphasized considering the 
    $r_s$-dependence of the screening wave vector Q$_{Exp}$=$\Omega_{pi}$/c$_{Exp}$, 
    shown in the insert of Fig.~\ref{fig4}.
    Here, $Q_{Exp}$ is compared with $Q_{RPA}$ and with the SSTL
    theoretical prediction ~\cite{SingwiPRB70}, which is an attempt
    to extend the range of the RPA by introducing a local field correction readjusted to an
external field. Once more, experiment and theories 
    are in good agreement up to $r_{sth}$, while a clear deviation 
    is observable above $r_{sth}$. We also note that the SSLT theory 
    not only does not improve RPA in describing $Q_{Exp}$, but seems to
    predict a departure from RPA opposite to the one experimentally measured.
    Finally, it is very interesting to note that for $r_s \ge r_{sth}$, 
    $Q_{Exp}$ is extremely well reproduced by an empirical $r_s^{-3/2}$ dependence.
   
    In conclusion, we presented IXS measurements in expanded liquid metals with high $r_s$
    allowing the determination of the high frequency sound velocity: these data, 
    along with literature data for metals with lower $r_s$, emphasize the 
    breakdown of existing electron gas theories above a threshold value
    $r_{sth}$. Most interestingly, the dielectric function, when the electron-electron 
    correlations become increasingly important, shows a departure from RPA 
    that follows a characteristic behavior which depends only on the 
    electron density and not on the specific details of the considered systems. 
    Future theories on the electronic dynamics must be compared with such a finding
    in order to test their ability to describe the deviations from the RPA. 

    We acknowledge A. Shukla for useful discussions, M. Di Michiel for his help 
    in the acquisition and treatment of the S(Q) measurements, C. Henriquet for 
    his contribution to the development of the experimental set-up, and C. Lapras 
    for technical support. C.A.B. was supported by the DOE under grant DE-FG02-99ER45772.


\begin{thebibliography}{24}
\expandafter\ifx\csname natexlab\endcsname\relax\def\natexlab#1{#1}\fi
\expandafter\ifx\csname bibnamefont\endcsname\relax
  \def\bibnamefont#1{#1}\fi
\expandafter\ifx\csname bibfnamefont\endcsname\relax
  \def\bibfnamefont#1{#1}\fi
\expandafter\ifx\csname citenamefont\endcsname\relax
  \def\citenamefont#1{#1}\fi
\expandafter\ifx\csname url\endcsname\relax
  \def\url#1{\texttt{#1}}\fi
\expandafter\ifx\csname urlprefix\endcsname\relax\def\urlprefix{URL }\fi
\providecommand{\bibinfo}[2]{#2}
\providecommand{\eprint}[2][]{\url{#2}}

\bibitem[{\citenamefont{Ashcroft and Mermin}(1976)}]{AshcroftMermin}
\bibinfo{author}{\bibfnamefont{N.~W.} \bibnamefont{Ashcroft}} \bibnamefont{and}
  \bibinfo{author}{\bibfnamefont{N.~D.} \bibnamefont{Mermin}},
  \emph{\bibinfo{title}{Solid State Physics}} (\bibinfo{publisher}{W. B.
  Saunders Company}, \bibinfo{year}{1976}).

\bibitem[{\citenamefont{Bove et~al.}(2001)\citenamefont{Bove, Sacchetti,
  Petrillo, Dorner, Formisano, and Barocchi}}]{BovePRL01}
\bibinfo{author}{\bibfnamefont{L.~E.} \bibnamefont{Bove}},
  \bibinfo{author}{\bibfnamefont{F.}~\bibnamefont{Sacchetti}},
  \bibinfo{author}{\bibfnamefont{C.}~\bibnamefont{Petrillo}},
  \bibinfo{author}{\bibfnamefont{B.}~\bibnamefont{Dorner}},
  \bibinfo{author}{\bibfnamefont{F.}~\bibnamefont{Formisano}},
  \bibnamefont{and} \bibinfo{author}{\bibfnamefont{F.}~\bibnamefont{Barocchi}},
  \bibinfo{journal}{Phys.\ Rev.\ Lett.} \textbf{\bibinfo{volume}{87}},
  \bibinfo{pages}{215504} (\bibinfo{year}{2001}).

\bibitem[{\citenamefont{Burns et~al.}(1999)\citenamefont{Burns, Abbamonte,
  Isaacs, and Platzman}}]{BurnsPRL99}
\bibinfo{author}{\bibfnamefont{C.~A.} \bibnamefont{Burns}},
  \bibinfo{author}{\bibfnamefont{P.}~\bibnamefont{Abbamonte}},
  \bibinfo{author}{\bibfnamefont{E.~D.} \bibnamefont{Isaacs}},
  \bibnamefont{and} \bibinfo{author}{\bibfnamefont{P.~M.}
  \bibnamefont{Platzman}}, \bibinfo{journal}{Phys.\ Rev.\ Lett.}
  \textbf{\bibinfo{volume}{83}}, \bibinfo{pages}{2390} (\bibinfo{year}{1999}).

\bibitem[{\citenamefont{Burns et~al.}(2002)\citenamefont{Burns, Giura, Said,
  Shukla, Vank\`{o}, Tuel-Benckendorf, Isaacs, and Platzman}}]{BurnsPRL02}
\bibinfo{author}{\bibfnamefont{C.~A.} \bibnamefont{Burns}},
  \bibinfo{author}{\bibfnamefont{P.}~\bibnamefont{Giura}},
  \bibinfo{author}{\bibfnamefont{A.}~\bibnamefont{Said}},
  \bibinfo{author}{\bibfnamefont{A.}~\bibnamefont{Shukla}},
  \bibinfo{author}{\bibfnamefont{G.}~\bibnamefont{Vank\`{o}}},
  \bibinfo{author}{\bibfnamefont{M.}~\bibnamefont{Tuel-Benckendorf}},
  \bibinfo{author}{\bibfnamefont{E.~D.} \bibnamefont{Isaacs}},
  \bibnamefont{and} \bibinfo{author}{\bibfnamefont{P.~M.}
  \bibnamefont{Platzman}}, \bibinfo{journal}{Phys.\ Rev.\ Lett.}
  \textbf{\bibinfo{volume}{89}}, \bibinfo{pages}{236404}
  (\bibinfo{year}{2002}).

\bibitem[{\citenamefont{Sacchetti et~al.}(2003)\citenamefont{Sacchetti,
  Guarini, Petrillo, Bove, Dorner, Demmel, and Barocchi}}]{SacchettiPRB03}
\bibinfo{author}{\bibfnamefont{F.}~\bibnamefont{Sacchetti}},
  \bibinfo{author}{\bibfnamefont{E.}~\bibnamefont{Guarini}},
  \bibinfo{author}{\bibfnamefont{C.}~\bibnamefont{Petrillo}},
  \bibinfo{author}{\bibfnamefont{L.~E.} \bibnamefont{Bove}},
  \bibinfo{author}{\bibfnamefont{B.}~\bibnamefont{Dorner}},
  \bibinfo{author}{\bibfnamefont{F.}~\bibnamefont{Demmel}}, \bibnamefont{and}
  \bibinfo{author}{\bibfnamefont{F.}~\bibnamefont{Barocchi}},
  \bibinfo{journal}{Phys.\ Rev.\ B} \textbf{\bibinfo{volume}{67}},
  \bibinfo{pages}{014207} (\bibinfo{year}{2003}).

\bibitem[{\citenamefont{Bove et~al.}(2003)\citenamefont{Bove, Dorner, Petrillo,
  Sacchetti, and Suck}}]{BovePRB03}
\bibinfo{author}{\bibfnamefont{L.~E.} \bibnamefont{Bove}},
  \bibinfo{author}{\bibfnamefont{B.}~\bibnamefont{Dorner}},
  \bibinfo{author}{\bibfnamefont{C.}~\bibnamefont{Petrillo}},
  \bibinfo{author}{\bibfnamefont{F.}~\bibnamefont{Sacchetti}},
  \bibnamefont{and} \bibinfo{author}{\bibfnamefont{J.~B.} \bibnamefont{Suck}},
  \bibinfo{journal}{Phys.\ Rev.\ B} \textbf{\bibinfo{volume}{68}},
  \bibinfo{pages}{024208} (\bibinfo{year}{2003}).

\bibitem[{\citenamefont{Copley and Rowe}(1974{\natexlab{a}})}]{CopleyPRA74}
\bibinfo{author}{\bibfnamefont{J.~R.~D.} \bibnamefont{Copley}}
  \bibnamefont{and} \bibinfo{author}{\bibfnamefont{J.~M.} \bibnamefont{Rowe}},
  \bibinfo{journal}{Phys.\ Rev.\ A} \textbf{\bibinfo{volume}{9}},
  \bibinfo{pages}{1656} (\bibinfo{year}{1974}{\natexlab{a}}).

\bibitem[{\citenamefont{Copley and Rowe}(1974{\natexlab{b}})}]{CopleyPRL74}
\bibinfo{author}{\bibfnamefont{J.~R.~D.} \bibnamefont{Copley}}
  \bibnamefont{and} \bibinfo{author}{\bibfnamefont{J.~M.} \bibnamefont{Rowe}},
  \bibinfo{journal}{Phys.\ Rev.\ Lett.} \textbf{\bibinfo{volume}{32}},
  \bibinfo{pages}{49} (\bibinfo{year}{1974}{\natexlab{b}}).

\bibitem[{\citenamefont{Bove et~al.}(2000)\citenamefont{Bove, Sacchetti,
  Petrillo, and Dorner}}]{BovePRL00}
\bibinfo{author}{\bibfnamefont{L.~E.} \bibnamefont{Bove}},
  \bibinfo{author}{\bibfnamefont{F.}~\bibnamefont{Sacchetti}},
  \bibinfo{author}{\bibfnamefont{C.}~\bibnamefont{Petrillo}}, \bibnamefont{and}
  \bibinfo{author}{\bibfnamefont{B.}~\bibnamefont{Dorner}},
  \bibinfo{journal}{Phys.\ Rev.\ Lett.} \textbf{\bibinfo{volume}{85}},
  \bibinfo{pages}{5352} (\bibinfo{year}{2000}).

\bibitem[{\citenamefont{Scopigno et~al.}(2002)\citenamefont{Scopigno, Balucani,
  Ruocco, and Sette}}]{ScopignoPRE02}
\bibinfo{author}{\bibfnamefont{T.}~\bibnamefont{Scopigno}},
  \bibinfo{author}{\bibfnamefont{U.}~\bibnamefont{Balucani}},
  \bibinfo{author}{\bibfnamefont{G.}~\bibnamefont{Ruocco}}, \bibnamefont{and}
  \bibinfo{author}{\bibfnamefont{F.}~\bibnamefont{Sette}},
  \bibinfo{journal}{Phys.\ Rev.\ E} \textbf{\bibinfo{volume}{65}},
  \bibinfo{pages}{031205} (\bibinfo{year}{2002}).

\bibitem[{\citenamefont{Scopigno
  et~al.}(2000{\natexlab{a}})\citenamefont{Scopigno, Balucani, Cunsolo,
  Masciovecchio, Ruocco, Sette, and Verbeni}}]{ScopignoEL00}
\bibinfo{author}{\bibfnamefont{T.}~\bibnamefont{Scopigno}},
  \bibinfo{author}{\bibfnamefont{U.}~\bibnamefont{Balucani}},
  \bibinfo{author}{\bibfnamefont{A.}~\bibnamefont{Cunsolo}},
  \bibinfo{author}{\bibfnamefont{C.}~\bibnamefont{Masciovecchio}},
  \bibinfo{author}{\bibfnamefont{G.}~\bibnamefont{Ruocco}},
  \bibinfo{author}{\bibfnamefont{F.}~\bibnamefont{Sette}}, \bibnamefont{and}
  \bibinfo{author}{\bibfnamefont{R.}~\bibnamefont{Verbeni}},
  \bibinfo{journal}{Europhys.\ Lett.} \textbf{\bibinfo{volume}{50}},
  \bibinfo{pages}{189} (\bibinfo{year}{2000}{\natexlab{a}}).

\bibitem[{\citenamefont{Soderstrom et~al.}(1980)\citenamefont{Soderstrom,
  Copley, Suck, and Dorner}}]{SoderstromJPF80}
\bibinfo{author}{\bibfnamefont{O.}~\bibnamefont{Soderstrom}},
  \bibinfo{author}{\bibfnamefont{J.~R.~D.} \bibnamefont{Copley}},
  \bibinfo{author}{\bibfnamefont{J.-B.} \bibnamefont{Suck}}, \bibnamefont{and}
  \bibinfo{author}{\bibfnamefont{B.}~\bibnamefont{Dorner}},
  \bibinfo{journal}{J.\ Phys.\ F} \textbf{\bibinfo{volume}{10}},
  \bibinfo{pages}{L151} (\bibinfo{year}{1980}).

\bibitem[{\citenamefont{Scopigno
  et~al.}(2000{\natexlab{b}})\citenamefont{Scopigno, Balucani, Ruocco, and
  Sette}}]{ScopignoPRE00}
\bibinfo{author}{\bibfnamefont{T.}~\bibnamefont{Scopigno}},
  \bibinfo{author}{\bibfnamefont{U.}~\bibnamefont{Balucani}},
  \bibinfo{author}{\bibfnamefont{G.}~\bibnamefont{Ruocco}}, \bibnamefont{and}
  \bibinfo{author}{\bibfnamefont{F.}~\bibnamefont{Sette}},
  \bibinfo{journal}{Phys.\ Rev.\ E} \textbf{\bibinfo{volume}{63}},
  \bibinfo{pages}{011210} (\bibinfo{year}{2000}{\natexlab{b}}).

\bibitem[{\citenamefont{Cabrillo et~al.}(2002)\citenamefont{Cabrillo, Bermejo,
  Alvarez, Verkerk, Maira-Vidal, Bennington, and Mart\`{i}n}}]{CabrilloPRL02}
\bibinfo{author}{\bibfnamefont{C.}~\bibnamefont{Cabrillo}},
  \bibinfo{author}{\bibfnamefont{F.~J.} \bibnamefont{Bermejo}},
  \bibinfo{author}{\bibfnamefont{M.}~\bibnamefont{Alvarez}},
  \bibinfo{author}{\bibfnamefont{P.}~\bibnamefont{Verkerk}},
  \bibinfo{author}{\bibfnamefont{A.}~\bibnamefont{Maira-Vidal}},
  \bibinfo{author}{\bibfnamefont{S.~M.} \bibnamefont{Bennington}},
  \bibnamefont{and}
  \bibinfo{author}{\bibfnamefont{D.}~\bibnamefont{Mart\`{i}n}},
  \bibinfo{journal}{Phys.\ Rev.\ Lett.} \textbf{\bibinfo{volume}{89}},
  \bibinfo{pages}{075508} (\bibinfo{year}{2002}).

\bibitem[{\citenamefont{Bodensteiner et~al.}(1992)\citenamefont{Bodensteiner,
  Morkel, Gl\"{a}ser, and Dorner}}]{DornerPRA92}
\bibinfo{author}{\bibfnamefont{T.}~\bibnamefont{Bodensteiner}},
  \bibinfo{author}{\bibfnamefont{C.}~\bibnamefont{Morkel}},
  \bibinfo{author}{\bibfnamefont{W.}~\bibnamefont{Gl\"{a}ser}},
  \bibnamefont{and} \bibinfo{author}{\bibfnamefont{B.}~\bibnamefont{Dorner}},
  \bibinfo{journal}{Phys.\ Rev.\ A} \textbf{\bibinfo{volume}{45}},
  \bibinfo{pages}{5709} (\bibinfo{year}{1992}).

\bibitem[{\citenamefont{Masciovecchio et~al.}(1996)\citenamefont{Masciovecchio,
  Bergmann, Krisch, Ruocco, Sette, and Verbeni}}]{MasciovecchioNIMPR96}
\bibinfo{author}{\bibfnamefont{C.}~\bibnamefont{Masciovecchio}},
  \bibinfo{author}{\bibfnamefont{U.}~\bibnamefont{Bergmann}},
  \bibinfo{author}{\bibfnamefont{M.}~\bibnamefont{Krisch}},
  \bibinfo{author}{\bibfnamefont{G.}~\bibnamefont{Ruocco}},
  \bibinfo{author}{\bibfnamefont{F.}~\bibnamefont{Sette}}, \bibnamefont{and}
  \bibinfo{author}{\bibfnamefont{R.}~\bibnamefont{Verbeni}},
  \bibinfo{journal}{Nucl.\ Instrum.\ Methods Phys.\ Res.}
  \textbf{\bibinfo{volume}{Sect B 111}}, \bibinfo{pages}{181}
  (\bibinfo{year}{1996}).

\bibitem[{\citenamefont{Balucani and Zoppi}(1994)}]{Balucani94}
\bibinfo{author}{\bibfnamefont{U.}~\bibnamefont{Balucani}} \bibnamefont{and}
  \bibinfo{author}{\bibfnamefont{M.}~\bibnamefont{Zoppi}},
  \emph{\bibinfo{title}{Dynamics of the Liquid State}}
  (\bibinfo{publisher}{Clarendon press Oxford}, \bibinfo{year}{1994}).

\bibitem[{\citenamefont{Giura et~al.}()\citenamefont{Giura, Angelini,
  Dimichiel, and Burns}}]{S(Q)nostre}
\bibinfo{author}{\bibfnamefont{P.}~\bibnamefont{Giura}},
  \bibinfo{author}{\bibfnamefont{R.}~\bibnamefont{Angelini}},
  \bibinfo{author}{\bibfnamefont{M.}~\bibnamefont{Dimichiel}},
  \bibnamefont{and} \bibinfo{author}{\bibfnamefont{C.}~\bibnamefont{Burns}},
  \bibinfo{note}{to be published.}

\bibitem[{\citenamefont{Burns et~al.}(2001)\citenamefont{Burns, Platzman, Sinn,
  Alatas, and Alp}}]{BurnsPRL01}
\bibinfo{author}{\bibfnamefont{C.~A.} \bibnamefont{Burns}},
  \bibinfo{author}{\bibfnamefont{P.~M.} \bibnamefont{Platzman}},
  \bibinfo{author}{\bibfnamefont{H.}~\bibnamefont{Sinn}},
  \bibinfo{author}{\bibfnamefont{A.}~\bibnamefont{Alatas}}, \bibnamefont{and}
  \bibinfo{author}{\bibfnamefont{E.~E.} \bibnamefont{Alp}},
  \bibinfo{journal}{Phys.\ Rev.\ Lett.} \textbf{\bibinfo{volume}{86}},
  \bibinfo{pages}{2357} (\bibinfo{year}{2001}).

\bibitem[{not()}]{nota1}
\bibinfo{note}{These values of $c_{Exp}$ are larger than the ultrasonic ones
  reported by Bowen et al. Phys. Rev. 168, 114 (1968). This has to be expected
  since, in these systems, the ultrasound technique probes the speed of sound
  in the viscous regime, as opposed to the elastic regime probed by inelastic
  X-ray and neutron scattering.}

\bibitem[{\citenamefont{Wasse et~al.}(1999)\citenamefont{Wasse, Hayama,
  Skipper, Benmore, and Soper}}]{WasseJCP99}
\bibinfo{author}{\bibfnamefont{J.~C.} \bibnamefont{Wasse}},
  \bibinfo{author}{\bibfnamefont{S.}~\bibnamefont{Hayama}},
  \bibinfo{author}{\bibfnamefont{N.~T.} \bibnamefont{Skipper}},
  \bibinfo{author}{\bibfnamefont{C.~J.} \bibnamefont{Benmore}},
  \bibnamefont{and} \bibinfo{author}{\bibfnamefont{A.~K.} \bibnamefont{Soper}},
  \bibinfo{journal}{J.\ Chem.\ Phys.} \textbf{\bibinfo{volume}{111}},
  \bibinfo{pages}{2028} (\bibinfo{year}{1999}).

\bibitem[{\citenamefont{Wasse et~al.}(2000)\citenamefont{Wasse, Hayama,
  Skipper, and Fischer}}]{WassePRB00}
\bibinfo{author}{\bibfnamefont{J.~C.} \bibnamefont{Wasse}},
  \bibinfo{author}{\bibfnamefont{S.}~\bibnamefont{Hayama}},
  \bibinfo{author}{\bibfnamefont{N.~T.} \bibnamefont{Skipper}},
  \bibnamefont{and} \bibinfo{author}{\bibfnamefont{H.~E.}
  \bibnamefont{Fischer}}, \bibinfo{journal}{Phys.\ Rev.\ B}
  \textbf{\bibinfo{volume}{61}}, \bibinfo{pages}{11993} (\bibinfo{year}{2000}).

\bibitem[{\citenamefont{Knapp and Bale}(1978)}]{KnappJAC78}
\bibinfo{author}{\bibfnamefont{D.~N.} \bibnamefont{Knapp}} \bibnamefont{and}
  \bibinfo{author}{\bibfnamefont{H.~D.} \bibnamefont{Bale}},
  \bibinfo{journal}{J.\ Appl.\ Crystallogr.} \textbf{\bibinfo{volume}{11}},
  \bibinfo{pages}{606} (\bibinfo{year}{1978}).

\bibitem[{\citenamefont{Singwi et~al.}(1970)\citenamefont{Singwi,
  Sj\"{o}lander, Tosi, and Land}}]{SingwiPRB70}
\bibinfo{author}{\bibfnamefont{K.~S.} \bibnamefont{Singwi}},
  \bibinfo{author}{\bibfnamefont{A.}~\bibnamefont{Sj\"{o}lander}},
  \bibinfo{author}{\bibfnamefont{M.~P.} \bibnamefont{Tosi}}, \bibnamefont{and}
  \bibinfo{author}{\bibfnamefont{R.~H.} \bibnamefont{Land}},
  \bibinfo{journal}{Phys.\ Rev.\ B} \textbf{\bibinfo{volume}{1}},
  \bibinfo{pages}{1044} (\bibinfo{year}{1970}).

\end{thebibliography}

    \end{document}